\documentclass[final,5p,times,twocolumn]{elsarticle}

\usepackage{amssymb}
\usepackage{hyperref}

\journal{Nuclear Physics B}

\begin{document}

\begin{frontmatter}



\title{Cherenkov Detectors in Astroparticle Physics}


\author{Christian Spiering}

\affiliation{organization={DESY}, 
            addressline={Platanenallee 6}, 
            city={Zeuthen},
            postcode={D-15738}, 
            country={Germany}}

\begin{abstract}
Cherenkov techniques are widely used in astroparticle experiments. This article reviews the various detection principles and the corresponding experiments, including some of the physics breakthroughs.  In particular, it traces the development since the mid of the 1990s, a period when the field took a particularly dynamic development. 
\end{abstract}


\end{frontmatter}


\section*{Introduction}
\label{intro}

When Pavel Cherenkov in 1934 discovered the radiation named after him \cite{Cherenkov}, no one could have imagined the enormous significance which this discovery would later have for particle and astroparticle physics. What concerns the latter, it took until 1953 that Bill Galbraith and John Jelley observed for the first time Cherenkov light produced by cosmic rays passing through the atmosphere \cite{Jelley}.  Seven years later, at the ICRC in 1959, Giuseppe Cocconi predicted that the Crab Nebula should be a strong emitter of gamma rays at TeV energies \cite{Cocconi} -- a key prediction for the field of astroparticle physics. This stimulated further work, most notably the construction of the first air-Cherenkov telescope by Alexandr Chudakov in the early 1960s. His telescope consisted of 12 mirrors of 1.5 m diameter, each focusing the light to a single photomultiplier. Observed sources included Cygnus-A and the Crab Nebula but, in the absence of a signal, Chudakov only could derive upper limits on the gamma-ray flux \cite{Chudakov}. Seen from today, this is no surprise: compared to the cosmic-ray background, the gamma-ray fluxes are much too small to be identified without using either imaging or timing techniques.  Another 25 years had to pass before the first cosmic source of TeV gamma rays could be pinpointed: the Crab Nebula, identified in 1989 with the Whipple Cherenkov Imaging Telescope in Arizona \cite{Weekes}. Two years earlier, however, an even more spectacular result in Cherenkov-light based astroparticle physics had been achieved, in this case not with Cherenkov emitted from atmospheric particle showers, but with Cherenkov light emitted by positrons in big water tanks: the detection of anti-neutrinos from the Supernova 1987A by the Kamiokande and IMB detectors \cite{Hirata,Bionta}.

This review will cover detectors using just these two media, air and water (or ice).  The detectors can be classified according to their location (underground, underwater, ground based) and according to the technique (ring imaging, imaging of air showers, timing techniques). The next table relates location, Cherenkov medium and detection technique for the different detector classes, here also including radio Cherenkov detection in ice \cite{RNO-G} and space detectors \cite{AMS} which will not be discussed in the following. Table 2 gives emission angle and intensity of Cherenkov light in air and water/ice.

\begin{table} [h] \footnotesize{Table\,1: Location, radiation medium and techniques for Cherenkov detectors in astroparticle physics}\\  \\
\footnotesize{
\begin{tabular}{|l|l|l|l|}
\hline 
 Location	& Cherenkov Medium &	Technique & Example \\
\hline
Underground & Ultrapure water	& Ring imaging &	Super-\\
& & & Kamiokande \\
\hline
Underwater/ice &	Natural water/ice  &	Timing  & \\
           & & optical & IceCube \\            
           & & radio	 & RNO-G  \\
           & & &  (Greenland) \\
\hline
Ground	& Atmosphere	&  Imaging &  	H.E.S.S. \\
\hline
		& Atmosphere & Timing	& TAIGA \\
	    & water in tanks	& Timing	 & HAWC \\
\hline
Space, balloons	& e.g. NaF	& Ring imaging	& AMS \\
\hline

\end{tabular}
}
\end{table}

\begin{table} [h] \footnotesize{Table\,2: Cherenkov emission angle and photon intensity (300\,nm$<\lambda<$600\,nm) for a single-charged particle moving with v/c$\sim1$ in air and in water}\\ \\
\begin{tabular}{|l|l|l|}
\hline
& Air & Water/Ice \\
\hline
Emission angle &	1.1° (1.4°) for &  41. 2°/ 40.3° \\
  &   8 (0) km altitude & \\
\hline 
Intensity & $\approx$15\,m &	$\approx 3\times10^4$\,m \\
\hline
\end{tabular}
\end{table}

Actually, I have given a similar talk, "Cherenkov imaging and timing techniques in astroparticle physics"  at the 1995 RICH conference in Uppsala \cite{Spiering-1995}.  I will therefore take the opportunity to compare the status of today with that from 1995. This will illustrate the bold progress in astroparticle physics over the last three to four decades, including the central role which Cherenkov detection techniques have played and are playing for this amazing development.

\section{RICH detectors underground}
\label{underground}

\begin{figure}[ht]
\includegraphics[width=0.53\columnwidth]{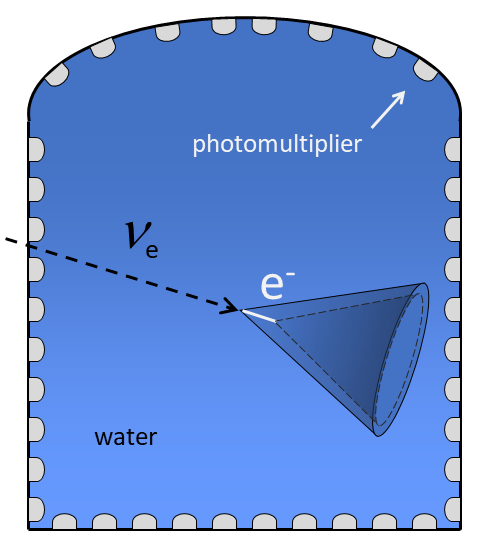}
\hspace{0.5cm}
\includegraphics[width=0.27\columnwidth]{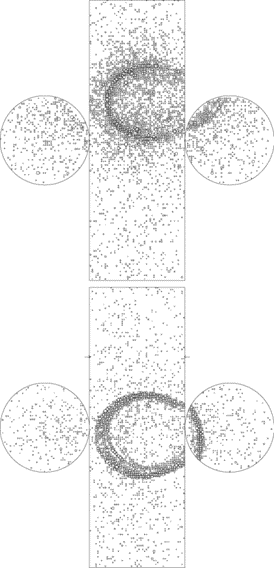}
\vspace{-0.2cm}
\caption{
Left: The principle of underground Cherenkov neutrino detectors. Stopping particles generate a ring pattern, through-going particles a filled circle. Right: ring pattern of a stopping electron (top) and muon (bottom) in Super-Kamiokande.
\label{principle}
}
\end{figure}

The principle of RICH detectors underground is illustrated in Fig.\ref{principle}. The walls of a tank filled with ultrapure water are paved with photomultipliers tubes (PMTs). The PMTs record the ring pattern of the Cherenkov light which has been emitted by charged particles generated in neutrino interactions. The larger the surface covered by PMTs, the lower the energy threshold for neutrino detection. Since electrons or positrons are easily showering up, their ring pattern is blurred, different to the pattern of muons. This allows distinguishing interactions of muon neutrinos from those of (anti-) electron neutrinos. The amount of light is a measure of the particle energy. The direction is mostly inferred from position and form of the ring pattern and can be improved by adding timing information.

Solar neutrinos are detected via elastic $\nu_e e^-$ scattering, where the final-state electron essentially preserves the neutrino direction. Supernova neutrinos are mostly visible via  
$\bar{\nu}_e + p \rightarrow  e^+ + n$ with poor directional information. Atmospheric (anti) neutrinos produce high-energy $e^{\pm}$ or $\mu^{\pm}$. The accessible energy ranges are 4-12 MeV (Sun), 5-40 MeV (Supernovae) and sub-GeV to TeV for atmospheric neutrinos. For the detection of the comparatively tiny fluxes of high-energy cosmic neutrinos, underground detectors turned out to be too small.

In 1995, the reference year chosen for my retrospect, two underground detectors already had written history: the Kamiokande detector in Japan and the IMB detector in the USA. Both devices had recorded neutrinos from Supernova 1987A (Kamiokande 12, IMB 8 events) \cite{Hirata,Bionta}, and Kamiokande, in addition, had measured solar neutrinos. In the case of Kamiokande, 1,000 20-inch PMTs observed a volume of 3\,ktons H$_2$O, in IMB 2,048 8-inch PMTs observed 9\,ktons of H$_2$O. IMB had been running between 1982 and 1991, while Kamiokande in 1995, after several upgrades, was in its last year of operation before Super-Kamiokande \cite{Super-K} took over. Super-K started operation in 1996.
Its tank is about 40\,m in diameter and 40\,m in height, filled with 50\,ktons of H$_2$O. The outermost 18\,ktons are used as veto layer. The inner volume is observed by about 11,000 20-inch PMTs (significantly improved in timing and charge resolution compared to Kamiokande), the veto layer by about 1,900 8-inch PMTs. 

Super-K has precisely measured neutrinos from the Sun and confirmed the solar neutrino deficit. Even more importantly, in 1998 it unambiguously measured oscillations of atmospheric neutrinos.

While the most obvious explanation of the solar neutrino deficit were flavor transitions in the Sun and not, for instance, neutrinos decaying on their way to Earth, the final confirmation of the first interpretation had to await results from the Sudbury Neutrino Observatory, SNO, in Canada \cite{SNO}. 
The core of SNO is an acrylic vessel which was filled with 1\,kton of heavy water (D$_2$O). This volume is observed by about 9,000 8-inch PMTs. It is surrounded by a veto layer filled with normal water. SNO took data between 1999 and 2006.  The deuteron target led to detectable reactions of all neutrino flavors and allowed to prove that – summing over all flavors – no deficit of solar neutrinos is observed. This result discarded all interpretations of the “solar neutrino puzzle” but the one of neutrino flavor transitions in the Sun. 

Two Nobel Prizes go to the account of Kamiokande (neutrinos from SN 1987A and from the Sun), and of Super-K and SNO (neutrino oscillations).  A fantastic record for underground Cherenkov detectors!

The water of Super-K has meanwhile been loaded with gadolinium in order to increase the neutron detection efficiency and thereby the sensitivity to the diffuse supernova neutrino flux. The SNO experiment is being followed by SNO+, using the same vessels and PMTs but employing a liquid-scintillator filling doped with Tellurium for the search for neutrino-less double beta decay. 

The big future underground water Cherenkov detector is Hyper-Kamiokande \cite{Kisiel}. It will contain 260 ktons of water, with a fiducial volume eight times that of Super-K. The inner detector is viewed by 40,000 20-inch PMTs which have twice the photon detection efficiency compared to Super-K PMTs, about two times better timing and charge resolution, and a lower dark current rate. The 20-inch PMTs will be supplemented by 1000 multi-PMT modules (19 3-inch PMTs in a pressure-tight housing) -- following a design invented by KM3NeT (see below). Hyper-Kamiokande construction began in early 2020, and the experiment is expected to start operations in 2027. The main goals of Hyper-K are the study of CP violation in the leptonic sector (using also an intense neutrino beam from J-PARC), the measurement of neutrino mixing parameters, the measurement of solar and supernova neutrinos, and the search for BSM phenomena like proton decay or dark matter.

\section*{Timing Detectors Underwater and in Ice}         
\label{underwater-detectors}

The primary goal of neutrino detectors in open water or ice (“neutrino telescopes”, NTs) is the detection of neutrinos from cosmic accelerators, at energies beyond those which are hopelessly dominated by atmospheric neutrinos, i.e. beyond several hundred GeV for point sources and a few tens of TeV for diffuse fluxes.  Because of the steeply falling flux of cosmic neutrinos ($E^{-2.2}$ to $E^{-2.5}$), this requires volumes on the cubic kilometer scale. Other phenomena addressed by NTs are the measurement of neutrino cross sections at energies beyond 10 TeV, Earth tomography, search for dark matter or magnetic monopoles, or environmental science. With densely equipped (sub)detectors, energies below 20 GeV are accessible, where oscillations of atmospheric neutrinos having crossed the Earth become visible. 

NTs consists of a matrix of optical modules (OMs) -- pressure tight glass spheres housing PMTs with their electronics. The OMs are arranged along strings deep in Oceans (ANTARES, KM3NeT), Lakes (Baikal GVD) or Antarctic ice (IceCube).   

\begin{figure}[ht]
\begin{center}
\includegraphics[width=1.0\columnwidth]{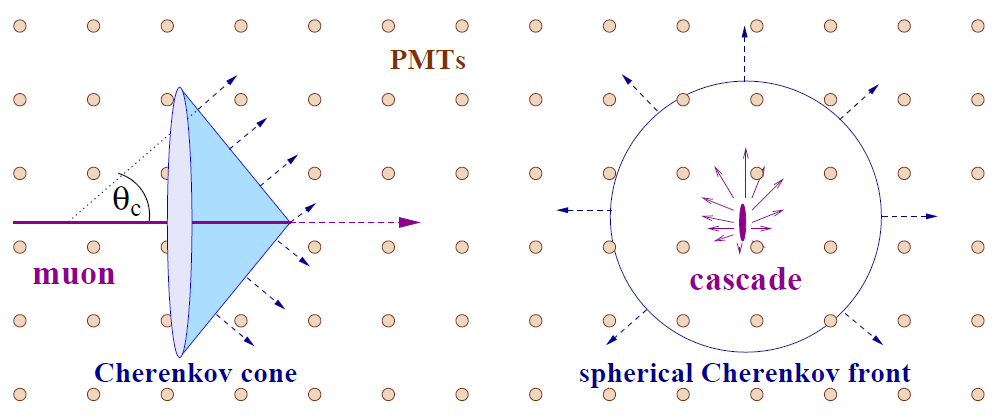}
\end{center}
\vspace{-0.2cm}
\caption{
Detection of muon tracks (left) and cascades (right) in underwater/ice detectors.
\label{underwater}
}
\end{figure}

Muons generated in charged current $\nu_{\mu}$ interactions leave a track generated by the Cherenkov light cone moving through the array (Fig.\ref{underwater}, left). The direction is inferred from the light arrival times at the PMTs, the deposited energy from the total detected amount of light. The elongated form with its long lever arm leads to a good angular resolution (IceCube $\approx 4^{\circ}$, KM3NeT $\approx 0.1^{\circ}$, at 100 TeV), while the muon energy can be only estimated via dE/dx to an order of magnitude (0.5 in $\log E_{\mu}$). Electrons and tauons generated in charged current $\nu_e$ and  $\nu_{\tau}$ interactions (as well as neutral current interactions) lead to a particle cascade of only 5-20 meters length and a more spherical light front (Fig.\ref{underwater},\,right). Naturally, the angular resolution is moderate (KM3NeT $\approx 2^{\circ}$ , IceCube  $\approx 10^{\circ}$, while the energy for these contained events can be determined  with 10$\%$ accuracy. The inferior angular resolution of IceCube is due to the stronger light scattering in ice.
 
Table\,3 gives an overview of the past, present and future projects in this field (here without the comparatively new projects P-ONE and TRIDENT).

\begin{table} [h] \footnotesize{Table\,3: Past, present and future NT projects. The milestone years give the times of first data taking with partial configurations, of detector completion (in brackets: 2022 status), and of project termination.}\\ \\
\footnotesize{
\begin{tabular}{|l|l|l|l|}
\hline

Experiment	&  Milestones	&  Location	&  Size (km$^3$) \\
\hline
NT200 & 1993/1998/2015	& Lake Baikal & 10$^{-4}$ \\
\hline
AMANDA	 & 1996/2000/2009 & South Pole	 & 0.015 \\
\hline
ANTARES & 2006/2008/2021	& Mediterr. Sea & 0.010 \\
\hline
\hline
IceCube	& 2004/2010/ - & South Pole	&  1.0 \\
\hline
GVD &  2015/2026/ - & Lake Baikal	 &  (0.5) 1.0 \\
\hline
KM3NeT/ARCA	& 2015/2027/ - & 	Mediterr. Sea & 1.2 \\
KM3NeT/ORCA & 	2017/2026/ - &	Mediterr. Sea & (0.001)\,0.007 \\
\hline
\hline
IceCube Gen2	 & 2028/2035/ - &  South Pole &  8 \\
\hline
KM3NeT Phase 3 &    2028/ - / - & 	Mediterr. Sea & $\approx$3 \\
\hline

\end{tabular}
}
\end{table}	

The pioneering project in the field was DUMAND (Deep Underwater Muon And Neutrino Detector) off the coast of Hawaii. Due to technical and financial problems, DUMAND was terminated in early 1995. At the time of RICH 1995, not a single clear neutrino event (e.g. an upward moving muon) had been detected underwater. NT200 in Lake Baikal had deployed only 36 of its final 192 OMs, and a clear neutrino identification from the data was still ahead. AMANDA at the South Pole was going to deploy its first four strings in the polar season 1995/1996 and would identify its first upward moving muons only in 1996 (together with NT200). See \cite{Spiering-2012} for a historical review.  

\begin{figure}[ht]
\begin{center}
\includegraphics[width=0.9\columnwidth]{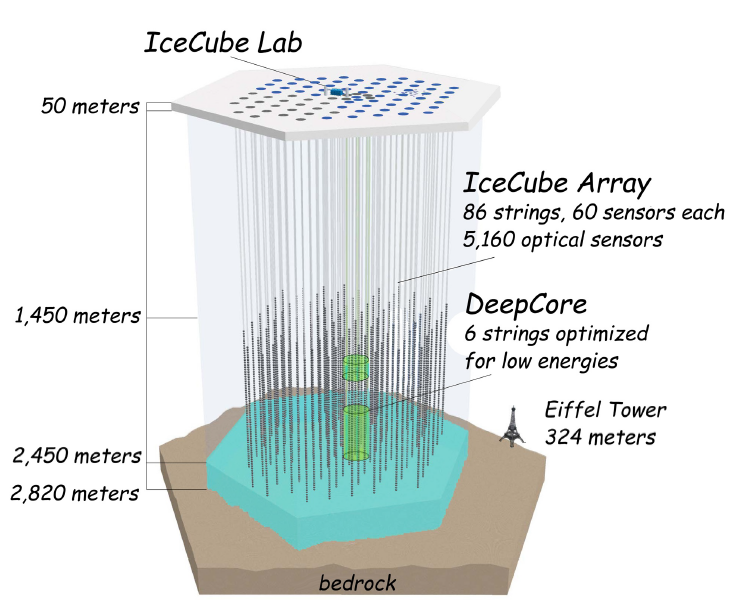}
\end{center}
\vspace{-0.3cm}
\caption{
 Schematic view of the IceCube Neutrino Observatory.
\label{icecube}
}
\end{figure}

The first-generation projects NT200, AMANDA and ANTARES have identified about 300, 7,000 and 10,000 upgoing tracks, respectively. Practically all these events could be attributed to interactions of atmospheric neutrinos. The field of high-energy neutrino astronomy was only opened by IceCube, the first NT on the cubic kilometer scale \cite{icecube}. Figure \ref{icecube} shows a schematic view of IceCube. It consists of 86 strings carrying altogether 5,160 downward-looking 10-inch PMTs covering a full cubic kilometer of ice at depths between 1450 m and 2450 m. A sub-volume called DeepCore is instrumented more densely than the rest, to detect neutrinos with energies down to 10 GeV. The IceTop Cherenkov surface array (frozen water in tanks with optical modules) serves for cosmic-ray studies and also provides some veto capability against muons and neutrinos from air showers.

IceCube has achieved three breakthrough results: The discovery of a diffuse flux of high-energy cosmic neutrinos in 2013 \cite{Aartsen-2013}, the first association of high-energy neutrinos to an astrophysical object, the blazar TXS 0506-056 (a variable source) in 2018 \cite{Aartsen-2018}, and the first clear identification of a steady astrophysical source, the active galaxy NGC 1068 in 2022 \cite{Abbasi-2022}. The discovery of the diffuse flux was meanwhile supported by data from ANTARES \cite{Albert-2018} and Baikal-GVD  \cite{GVD-2022} -- although with much lower significance (IceCube meanwhile more than $10\sigma$, ANTARES 1.6$\sigma$ and GVD 3.3$\sigma$). A fourth important result are the DeepCore constraints on neutrino oscillations which are similarly tight as those from accelerator experiments or Super-Kamiokande.
 
In the field season 2025/26, seven additional strings with newly developed optical modules and calibration devices will be added to the Deep Core region (“IceCube Upgrade”).  The main objectives of the upgrade are $a)$ a better understanding of the ice properties and a corresponding reduction of the systematic uncertainties, $b)$ sensitivity to neutrinos in the few-GeV range and a more precise measurement of oscillation phenomena and $c)$  the test of new hardware developments for a future array on the 8-cubic kilometer scale called IceCube-Gen2.

Two other projects on the cubic kilometer scale are currently under construction, Baikal-GVD and KM3NeT.

\begin{figure}[ht]
\begin{center}
\includegraphics[width=0.9\columnwidth]{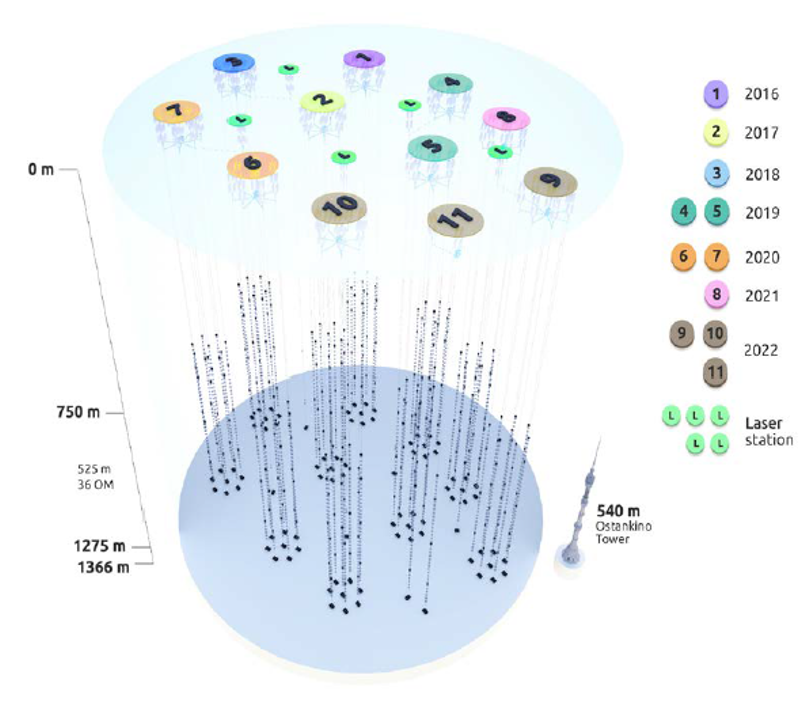}
\end{center}
\vspace{-0.3cm}
\caption{
 Schematic view of the presently deployed 10 clusters of GVD. Item 11 in the figure marks two isolated strings for tests of future technologies.
\label{gvd}
}
\end{figure}

Baikal-GVD (Gigaton Volume Detector) \cite{baikalgvd} will consist of eighteen 8-string clusters with a diameter of 120 m and a height of 525 m. At present, 10 of the 18 clusters are already deployed (see Fig.\,\ref{gvd}), completion of GVD is foreseen for 2026.  Each string carries 36 optical modules equipped with 10-inch, downward-looking PMTs.  First results include strong indications for a diffuse flux of neutrinos, as mentioned above.

KM3NeT \cite{km3net, Drakopoulou-RICH} will consist of building blocks of 115 strings each, with 18 OMs per string. KM3NeT aims at two separate detectors: ARCA (for Astroparticle Research with Cosmics in the Abyss) will consist of two building blocks for neutrino astronomy, with vertical distances between OMs of 36 m and a lateral distance between adjacent strings of 90 m;  ORCA (for Oscillation Research with Cosmics in the Abyss) will be one block for the measurement of the neutrino mass hierarchy, with vertical distances between OMs of 9 m and a lateral distance between adjacent strings of about 20 m. The volume of one ARCA block is about 0.6 km$^3$ and that of ORCA 0.007 km$^3$. The installation of ARCA near Capo Passero, East of Sicily (depth 3440 m) and of ORCA near Toulon (depth 2450 m) is ongoing. As of December 2022, 21 strings have been deployed at the ARCA site and 15 at the ORCA site. Completion of the full ARCA (ORCA) arrays is planned for 2027 (2026). 

\begin{figure}[ht]
\begin{center}
\includegraphics[width=0.5\columnwidth]{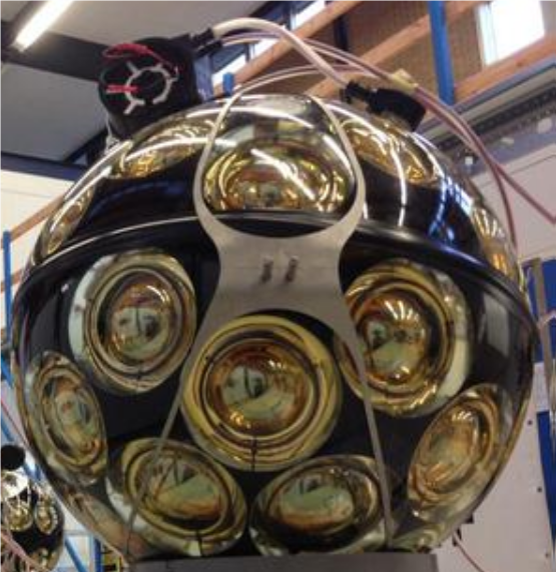}
\end{center}
\vspace{-0.3cm}
\caption{
 The KM3NeT digital optical module.
\label{km3OM}
}
\end{figure}

One of the stunning inventions in KM3NeT is its Optical Module (see Fig.\,\ref{km3OM}). Instead of one single 10-inch PMT it houses 31 3-inch PMTs. What at a first glance looks as just complicating the straightforward one-PMT design has several significant advantages: a three times higher photocathode area per OM, almost 4$\pi$ solid angle coverage, a better 1-vs-2 photo-electron separation and intrinsic directional information.

\begin{figure}[ht]
\begin{center}
\includegraphics[width=0.8\columnwidth]{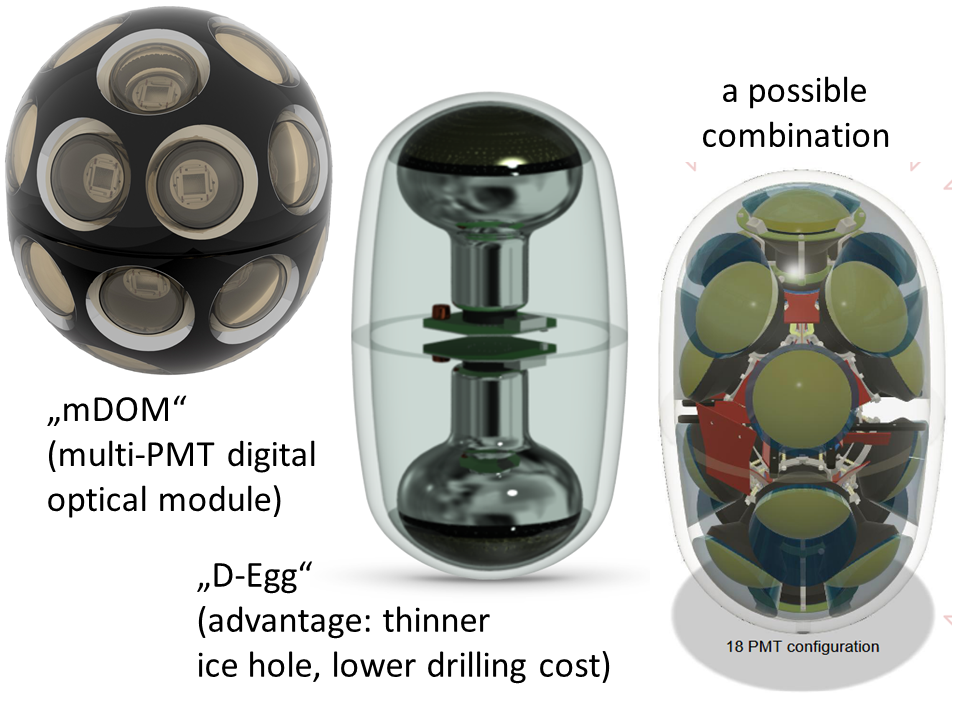}
\end{center}
\vspace{-0.3cm}
\caption{
 Optical Module options for IceCube-Gen2.
\label{Gen2OM}
}
\end{figure}

For IceCube’s follow-up project IceCube-Gen2 \cite{Gen2, Xu-RICH}, two baseline OM options are considered: the mDOM  (similar to the KM3NeT OM but with 24 3-inch PMTs) and a solution with two 8-inch PMTs, the “D-Egg” (see Fig.\,\ref{Gen2OM}). D-Egg would allow for thinner ice holes and therefore significantly reduced drilling cost. Other options like the one on the right side also are under investigation.

IceCube-Gen2 will comprise 120 strings spaced by 250 m instead of the current 125 meters in IceCube. This allows instrumenting an eightfold larger volume at only moderately increasing cost. The 8 km$^3$ optical array is envisaged to be extended by a radio array near the surface, covering an area of about 500 square kilometers. It will record the coherent Cherenkov radio emission of particle cascades at energies above a few tens of PeV and improve the sensitivity at the highest energies by almost two orders of magnitude.

There are two new projects, the one currently in an advanced prototype phase and the other one in an early conceptual phase. The first is named P-ONE (Pacific Ocean Neutrino Explorer \cite{p-one}). It is conceived as a multi-cluster array (similar to Baikal-GVD) and will be deployed at the West Coast of Canada, using the infrastructure of the Canadian Ocean Network ONC. The other, named TRIDENT, is a plan for a 7.5 km$^3$ NT in the South-China Sea \cite{trident}. First environmental results from an exploratory ship cruise are available.

Taken all together, the field of high-energy neutrino astronomy has made a giant leap since 1995: from embryonic configurations of NTs in Lake Baikal and at the South Pole to a cubic kilometer NT completed 12 years ago. Right now, two NTs of similar size are under construction and two others conceived. IceCube has led to several breakthrough results, proving that first steps into high-energy neutrino astronomy are possible with a cubic-kilometer NT. It can be taken for almost granted that a further order of magnitude in size will allow mapping the landscape of celestial high-energy neutrino sources -- both in terms of the number of sources and in terms of their character.

\section*{Imaging Air Cherenkov Telescopes (IACTs)}
\label{iact}

Air showers from gamma rays can be detected on ground by shower imaging or by timing (wave-front sampling) techniques.  History and development of these techniques are comprehensively described in a recent review \cite{bose-2022}. The present section is devoted to imaging techniques, the next section to timing techniques.

Imaging Air Cherenkov Telescopes  record the image in Cherenkov light of air showers generated by gamma rays in the atmosphere.  At energies above $\approx$20 GeV, gamma rays initiate electromagnetic cascades extending over several kilometers, with a maximum at a height of 10-15 km above sea level. The electrons and positrons in the cascade generate Cherenkov light. Its amount is proportional to the integrated track length of all particles and is therefore, to a good approximation, proportional to the initial gamma-ray energy. Due to the small emission angle of Cherenkov light in air, the Cherenkov light pool at ground level has a radius of only 100–150 m.

\begin{figure}[ht]
\begin{center}
\includegraphics[width=0.8\columnwidth]{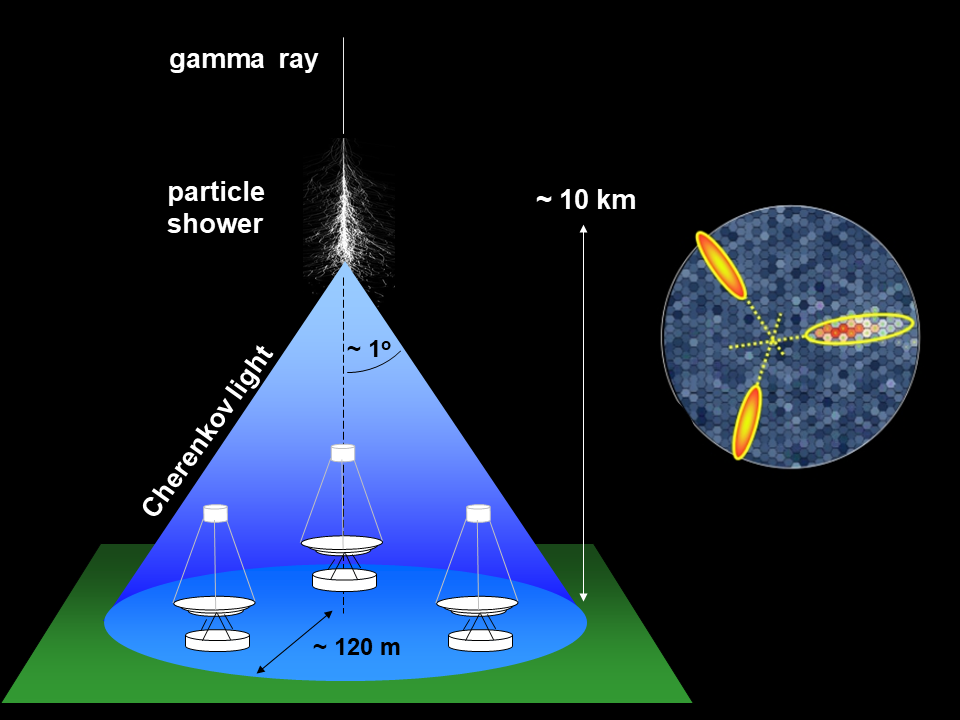}
\end{center}
\vspace{-0.3cm}
\caption{
Principle of Imaging Air Shower Telescopes.
\label{IACT}
}
\end{figure}

An IACT consists of a large segmented mirror which focuses the light to a matrix of PMTs which record the cigar-shaped image of the air shower as shown in Fig.\ref{IACT}.  The gamma-ray direction and energy are reconstructed from the recorded light pattern and intensity. Cascades induced by charged cosmic rays are three orders of magnitude more frequent than gamma rays. However, since their image is wider and fuzzier than that of gamma-ray showers, they can be efficiently suppressed – the finer the pixelization of the camera the more efficiently. Using more than one telescope (see Fig.\ref{IACT}) allows stereoscopic imaging and results in better angular and energy resolution as well as better background suppression.

The main parameters defining the quality of a telescope are the pixelization of the camera (the finer, the better are angular resolution and background suppression via image topology), the mirror size (the larger, the lower is the energy threshold), the altitude (the higher, the lower is the energy threshold), the quality of the night sky (low light pollution and good air quality) and the field of view.

\begin{figure}[ht]
\begin{center}
\includegraphics[width=0.7\columnwidth]{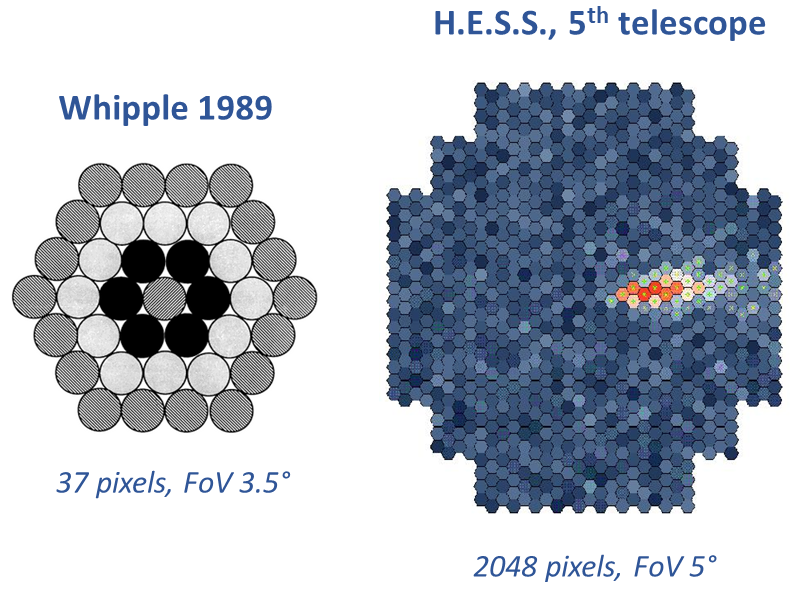}
\end{center}
\vspace{-0.4cm}
\caption{
Comparison of the 1989 Whipple camera and the camera of the fifth H.E.S.S. telescope (2012). Shown sizes are chosen according to the field of view (FoV).
\label{cameras}
}
\end{figure}

The Whipple Telescope at Mt. Hopkins in Arizona has pioneered the imaging technique by operating in the late 1980s an array of only 37 PMTs in the focal plane of a 10 m diameter mirror \cite{Weekes}. The limited resolution of only 37 pixels did not allow image analyses as used today (see Fig.\ref{cameras} for a comparison of the Whipple camera to a modern camera). Instead, the image was analyzed in terms of a simple but ingenious parametrization \cite{Hillas}. So, the collaboration could report in 1989 the first clear observation of a TeV gamma-ray source, the Crab Nebula, with a significance of 9$\sigma$.

In 1996, three TeV gamma-ray sources had been detected with the Whipple telescope: the Crab Nebula and two active galaxies, Mk 421 and Mk 501 (for the latter, the HEGRA telescope array on the Canary Island La Palma followed in 1997 \cite{bradbury-1997}). HEGRA was the first project using the stereoscopic technique which is also used by the present IACT working horses:  H.E.S.S. \cite{hess} in Namibia (5 telescopes), MAGIC \cite{magic}  on La Palma (2 telescopes) and VERITAS \cite{veritas}  in Arizona (4 telescopes). Table\,4 summarizes the basic parameters of these observatories. Figure \ref{cameras} illustrates the huge step in pixelization and corresponding shower image resolution made from the 1989 Whipple camera to the largest of the H.E.S.S. cameras.

\begin{table} [h] \footnotesize{Table\,4: Basic parameters of H.E.S.S., MAGIC and VERITAS. The fifth, large H.E.S.S. telescope started operation 9 years later than the other four, and also the second MAGIC telescope came only five years after the first one into operation} \\ \\
\begin{tabular}{|l|l|l|l|}
\hline
 & H.E.S.S.	& MAGIC & VERITAS \\
\hline
Altitude	& 1800 m	& 2200 m	& 1270 m \\
\hline
Dish diameter	& 4$\times$ 12 m & 2$\times$ 17 m & 	4$\times$12 m \\
& 1$\times$ 28 m &  & \\
\hline 
Nb. of pixels &	 4$\times$ 960	& 2$\times$ 576	& 4$\times$ 499 \\
& 1$\times$ 2048 & & \\
\hline
Field of view	& 5$^\circ$	 & 3.5$^\circ$ 	& 3.5$^\circ$ \\
\hline
Start of operation	& 2003/2012 & 2004/2009 & 2007 \\
\hline        
\end{tabular}
\end{table}   

\begin{figure}[ht]
\begin{center}
\includegraphics[width=1.0\columnwidth]{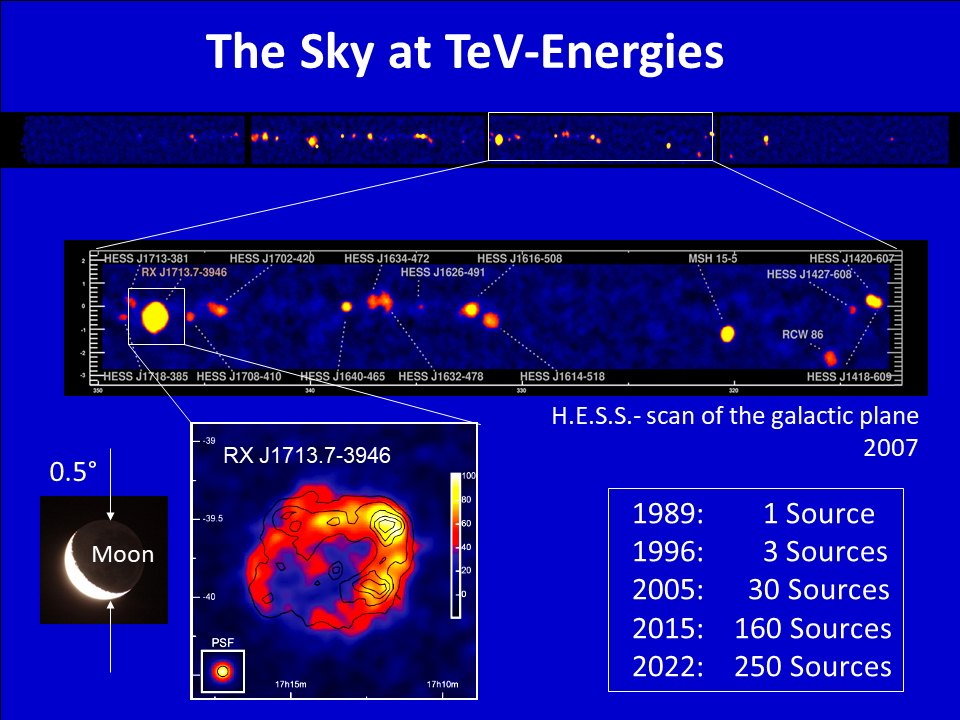}
\end{center}
\vspace{-0.3cm}
\caption{
Results of the second scan of the galactic plane in 2007. The rise in the number of detected sources is shown on the right side.
\label{scan}
}
\end{figure}

The field took an amazing development: from one source in 1989 to three sources in 1997, about thirty in 2005 and about 250 as of today \cite{bose-2022}. Figure \ref{scan} illustrates this development and the status in 2007, when H.E.S.S. published its second galactic scan \cite{voelk-2009}. Meanwhile, gamma-ray astronomy with IACTs is approaching standard astronomy in several aspects. Source positions can be determined with arc-second accuracy. The morphology of extended sources can be resolved on the arc-minute level. Variability or periodicity have been measured for time scales ranging from milliseconds to years. In addition, the gamma-ray spectrum can be measured over several decades in energy, from MeV (with satellites) to about one PeV (with earth-bound detectors).

\begin{figure}[ht]
\begin{center}
\includegraphics[width=1.0\columnwidth]{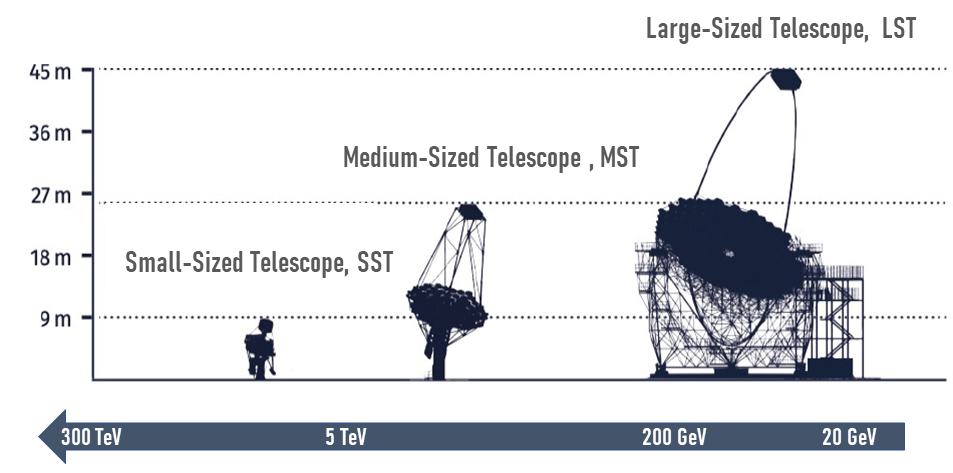}
\end{center}
\vspace{-0.2cm}
\caption{
 The three telescope types in CTA and the energy ranges for which they are optimized.
\label{cta}
}
\end{figure}

The next big step in the field is the Cherenkov Telescope Array, CTA \cite{cta,cta-RICH}. It will be installed at two sites, one in the Northern hemisphere on La Palma and the other in the Southern hemisphere close to the ESO Paranal Observatory in Chile. CTA will comprise telescopes of three sizes, LSTs (large size telescopes), MSTs (medium) and SSTs (small), which are focusing to different energy ranges (see Fig.\ref{cta}). Table 5 summarizes the basic parameters of these telescopes. The SSTs are equipped with SiPMs rather than PMTs, a technique which has been pioneered with the FACT telescope on La Palma \cite{fact}. SiPMs can take very high rates, enabling operation during full moon nights.

\begin{table} [h] \footnotesize {Table\,5: Basic parameters of the three telescope types in CTA. Two camera designs exist for the MST, one to be installed at the Southern site, the other at the Northern site.} \\ \\
\begin{tabular}{|l|l|l|l|}
\hline
& SST &	MST	& LST \\
\hline
Optics	& 2-mirror & 1-mirror & 1-mirror \\
 &  (Schwarzschild- & (Davis-Cotton)  &  (parabolic) \\
  & Couder) & & \\
\hline
Mirror diameter	& 4.3 m	&  12 m	&  23 m \\
\hline
Nb. of camera  &	2048   &	1855/1764 	& 1855  \\
pixels &  (SiPM) &  (PMT) &  (PMT) \\
\hline
Field of view	& 8.8$^\circ$	& 7$^\circ$  &	4.5$^\circ$ \\
\hline
\end{tabular}
\end{table}

According to the current plans (Dec. 2022), the Northern array will include 13 telescopes distributed over an area of about 0.5 km$^2$: four LSTs and nine MSTs. The array, which is optimized for the energy range 20 GeV to 5 TeV, will specialize in extragalactic sources, (gamma rays with much higher energies are absorbed by CMB over larger distances).  The Southern array will include 51 telescopes over a $\approx$3 km$^2$ area, consisting of 14 MSTs and 37 SSTs. This telescope configuration allows the southern array to focus on Galactic targets, optimizing its capabilities on the medium- and high-energy range (150 GeV --300 TeV). This so-called Alpha Configuration does not consider LSTs in the South, but it includes the preparation of the foundation for four of them, as well as the foundation for three more SSTs, to allow for the construction of these telescopes in a future enhancement of the array. There exists also an additional design for the MSTs, using the 2-mirror Schwarzschild-Couder option \cite{venere-RICH}. Its implementation in CTA depends on funding.  

\begin{figure}[ht]
\begin{center}
\includegraphics[width=1.0\columnwidth]{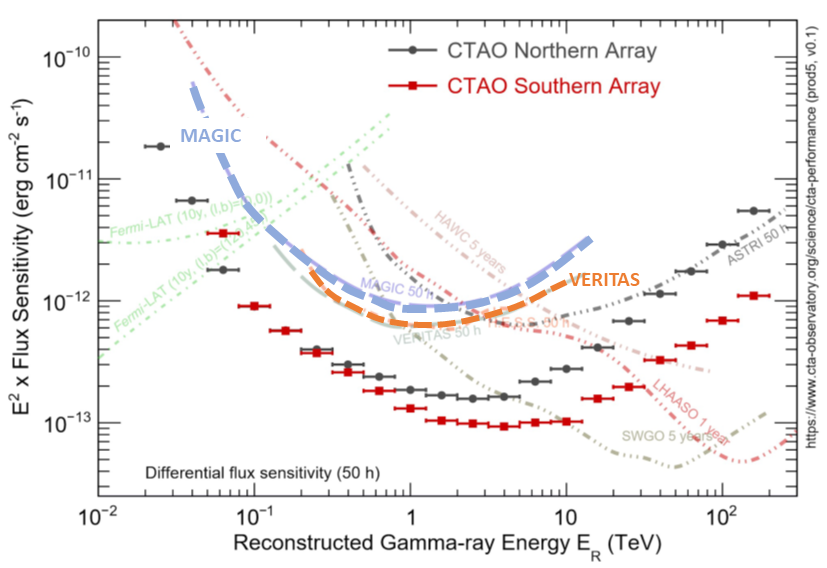}
\end{center}
\vspace{-0.4cm}
\caption{
Differential flux sensitivity of the current (H.E.S.S., MAGIC, VERITAS) and the future (CTA) ground-based IACTs. Also shown are the corresponding sensitivities for timing arrays like HAWC, LHAASO and SWGO (see the next section for more details on these instruments). The green, dash-dotted lines indicate the sensitivity of the satellite instrument Fermi-LAT for two different directions of observation. ASTRI is an array of two-mirror telescopes based on an early version of the CTA SST \cite{scuderi-RICH}.  Picture taken from \cite{cta}.
\label{sensitivity}
}
\end{figure}

Figure \ref{sensitivity} demonstrates the leap in sensitivity which CTA will achieve, compared to the present-generation IACTs.

\section*{Timing and Hybrid Detectors for Air Showers}
\label{TIMING}

\begin{figure}[ht]
\begin{center}
\includegraphics[width=0.8\columnwidth]{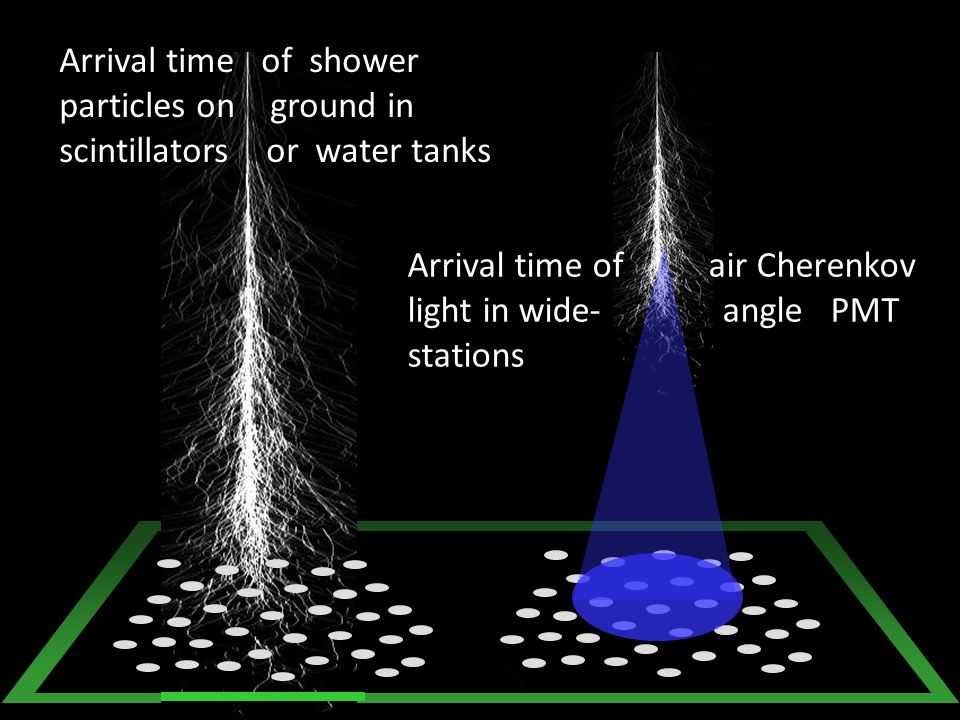}
\end{center}
\vspace{-0.2cm}
\caption{
Two methods of ground-based air shower detection with timing arrays.
\label{timing}
}
\end{figure}

Timing arrays record the arrival times of shower particles or of the Cherenkovlight from shower particles at ground (see Fig.\ref{timing}). One can then fit the arrival times to a conical front model and determine the main axis of the shower (impact point and angle). Furthermore, one can extract the energy from the amplitude information. The lateral distribution of particles or light helps distinguishing gamma-ray and hadron showers.

At altitudes exceeding 3-4 km above sea level, gamma-ray and cosmic-ray showers down to a few hundred GeV can reach the ground. They can be observed with arrays of scintillation or Cherenkov detectors, e.g. water tanks in which charged particles generate Cherenkov light (Fig.\ref{timing}, left).  Leptonic and hadronic showers can be separated according to event topologies and muon content (using often special counters at shallow depth to identify the more penetrating muons).  Table 6 lists a few prominent examples.

\begin{table} [h] \footnotesize{Table\,6: Examples for present air shower detectors recording the Cherenkov light generated by charged particles in water tanks. The two values for the HAWC area refer to the core detector and the full detector including the low-density outrigger array. LHAASO comprises not only water tank Cherenkov detectors but also scintillation and air Cherenkov detectors. The given threshold refers to the water tank detectors. CR = charged cosmic rays.} \\ \\
\begin{tabular}{|l|l|l|l|l|}
\hline
  &  Pierre Auger & IceTop &  HAWC  & LHAASO \\
\hline
Location & Chile  & South Pole & Mexico & Tibet \\
\hline
Altitude & 1.4\,km 	& 2.9\,km  & 4.1\,km  & 4.4\,km  \\
\hline
Area &  3000\,km$^2$  &	1\,km$^2$ & 0.02/ & 1\,km$^2$ \\
& & & 0.05\,km$^2$ & \\
\hline
Primary  & CR \hspace{2mm} ($\gamma,\nu$) & CR & CR, $\gamma$ &  CR, $\gamma$ \\
\hline
Energy & & & & \\
threshold & 10$^{17.5}$\,eV & 10$^{14}$eV & 10$^{11}$\,eV & 10$^{11}$\,eV \\
\hline
\end{tabular}
\end{table}

The most prominent detector of this class is the Pierre Auger Observatory with its 1660 water tanks, each 3.6 m in diameter. Auger is designed to detect charged cosmic rays at energies above 10$^{17.5}$ eV. 

The currently leading detector water tank detector with gamma-ray capabilities is HAWC \cite{hawc} in Mexico, at an altitude of 4100 m. HAWC consists of 300 water tanks covering 0.05\,km$^2$, each filled with 200 tons of water observed by 4 PMTs.  Gamma-ray showers tend to have a footprint which decreases steeply with the distance from the center of the shower. In contrast, the footprints of cosmic-ray showers are relatively messy and will appear "blotchy" when observed in the pattern of triggered PMTs. HAWC reaches an angular resolution of about 0.1$^{\circ}$ for gamma-ray energies larger than 10\,TeV.  For energies below 100\,TeV, HAWC is less sensitive than CTA. However, its large field of view and its high duty cycle (continuously instead of only during moonless nights) gives HAWC an advantage with respect to full-sky exploration and to steady gamma-ray sources. Moreover, its large field of view makes it particularly well suited to observe gamma-ray emission from extended objects.
  
A larger version of HAWC is discussed for the Southern hemisphere: SWGO (Southern Wide-field Gamma-ray Observatory \cite{swgo}. It is conceived to consist of an 80,000\,m$^2$ inner core with 80$\%$ coverage of water tanks, embedded in a 200,000\,m$^2$ outer region with 8$\%$ coverage. 

\begin{figure}[ht]
\begin{center}
\includegraphics[width=1.0\columnwidth]{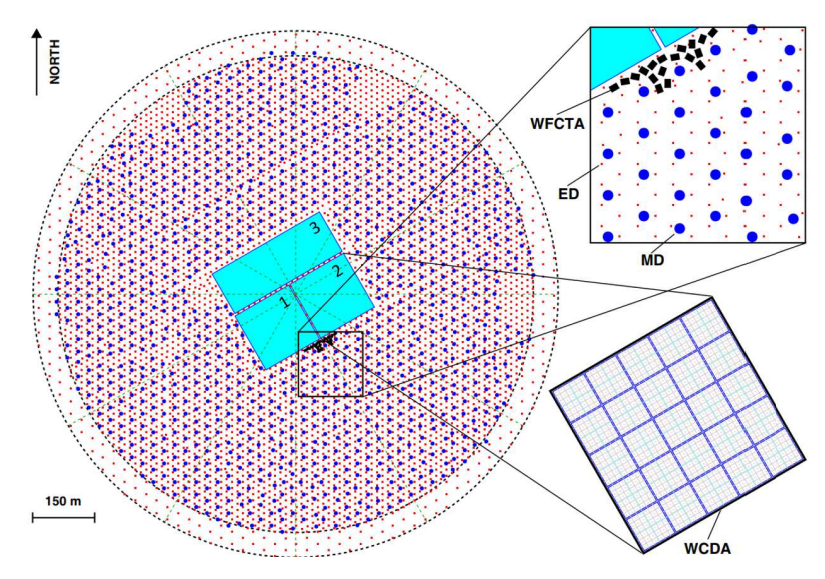}
\end{center}
\vspace{-0.1cm}
\caption{
The LHAASO detector array \cite{LHAASO-2021} (see text for explanations).
\label{lhaaso}
}
\end{figure}

Exciting results have recently reported from the LHAASO detector in Tibet \cite{LHAASO-2021}. This is a hybrid detector, including water tanks, scintillation detectors and wide-angle air Cherenkov detectors. The central part covers a 78,000\,m$^2$ area and is closely packed with water tanks (LHAASO-WCDA). Additional components are arrays of 1\,m$^2$ surface scintillation detectors (ED) and 32\,m$^2$ underground water Cherenkov tanks (MD). ED serves for detection of the electromagnetic part of the shower, MD for the detection of muons. The assembly is completed by a small array of Wide-Field air Cherenkov detectors (WFCTA). Figure \ref{lhaaso} gives a schematic top view of the array.

The initial LHAASO results seem heralding a new era in gamma-ray astronomy. With less than half of the components installed, first results include the discovery (larger $7\sigma$) of 12 gamma-ray sources with emission above 100\,TeV, the first “PeVatrons”, as well as a detailed analysis of the Crab Nebula spectrum, the latter including also information of WCDA \cite{LHAASO-2022}. LHAASO and the future SWGO extend the sensitivity of CTA towards highest energies (see Fig.\ref{sensitivity}).

The technique depicted in Figure \ref{timing} (right) was pioneered in the early 1990s by AIROBICC, a small experiment on the Canary Island La Palma. It consisted of a $7 \times 7$ matrix of wide-angle Cherenkov counters equipped each with one large PMT which measured the arrival time of the Cherenkov light front \cite{airobicc}. A present experiment based both on timing and imaging techniques is TAIGA (Tunka Advanced Instrument for cosmic ray physics and Gamma Astronomy) in the Siberian Tunka valley close to Lake Baikal. When finished, it will consist of 120 wide angle timing detectors spread over 1\,km$^2$ \cite{hiscore}, at least three imaging telescopes, and a large number of buried scintillation muon counters. The timing detector HiSCORE follows the same detection principle as AIROBICC and as Tunka-133 (a cosmic-ray detector at the same Siberian site \cite{tunka}). TAIGA has a much better time resolution, i.e. directional precision, than Tunka-133 and will allow good gamma/hadron separation at high energies. The imaging telescopes yield a superior directional resolution and improve, together with the muon counters, gamma/hadron separation.  With the stereoscopic operation of the first two IACTs, first gamma-ray sources have been identified up to $\approx$50 TeV energy, adding the HiSCORE data even up to energies of up to $\approx$100 TeV \cite{taiga}. 

\section*{Summary}
\label{summary}

Cherenkov techniques are essential tools of astroparticle physics. Enormous progress and several breakthrough results have been obtained during the past 25 years, for instance the confirmation of neutrino oscillations with the help of solar and atmospheric neutrinos. The realm of gamma-ray astronomy has been extended far into the TeV range. The number of identified TeV gamma-ray sources has increased by a factor of 100, including the detection of first PeV gamma-ray sources. The improved angular resolution of IACTs even allows revealing the morphology of sources. Last but not least, the sensitivity of neutrino telescopes has been improved by almost three orders of magnitude, and the window to the high-energy neutrino sky has been opened with the detection of a diffuse flux and of point sources of energetic neutrinos.

This incredible progress has been achieved due to several factors:
\begin{itemize} 
\item the size of the detectors, 
\vspace{-0.2cm}
\item advances in technology,
\vspace{-0.2cm}
\item sophisticated analysis methods,
\vspace{-0.2cm}
\item the choice of appropriate sites,
\vspace{-0.2cm}
\item the combination of complementary detection methods,
\vspace{-0.2cm}
\item the combination of information from different messengers (multi-messenger approach).
\end{itemize}

Next-generation projects like Hyper-Kamiokande, KM3NeT, Baikal-GVD, IceCube-Gen2, CTA, LHAASO or SWGO will continue this success story.

\section*{Acknowledgments}

I thank the organizers for their invitation to the RICH 2022 workshop, and Uli Katz and Johannes Knapp for helpful comments on the text.

\end{document}